%% file: paper.tex
\documentclass[conference]{IEEEtran}
\IEEEoverridecommandlockouts
\usepackage{subcaption}
\usepackage{tikz}
\usetikzlibrary{fit}

\newcommand{\panelbboxtop}{%
  \path[use as bounding box] (-0.8,-0.7) rectangle (5.2,2.35);
}
\newcommand{\paneltagtop}[1]{%
  \node[anchor=north west,font=\bfseries\small,inner sep=1pt, draw=none]
  at (-0.75,2.25) {#1};
}

\newcommand{\panelbboxbottom}{%
  \path[use as bounding box] (-0.8,-0.7) rectangle (5.2,3.4);
}
\newcommand{\paneltagbottom}[1]{%
  \node[anchor=north west,font=\bfseries\small,inner sep=1pt, draw=none]
  at (-0.75,3.25) {#1};
}

\newenvironment{smallblock}
  {%
   \par
   \begingroup
   \small
   \setlength{\parskip}{0pt}%
   \setlength{\abovedisplayskip}{3pt plus 1pt minus 1pt}%
   \setlength{\belowdisplayskip}{3pt plus 1pt minus 1pt}%
   \setlength{\abovedisplayshortskip}{0pt}%
   \setlength{\belowdisplayshortskip}{3pt plus 1pt minus 1pt}%
   \noindent
  }
  {%
   \par
   \endgroup
  }

\usepackage{cite}
\usepackage{amsmath,amssymb,amsfonts}
\usepackage{algorithmic}
\usepackage{graphicx}
\usepackage{subcaption}
\usepackage{textcomp}
\usepackage{xcolor}
\usepackage{pgfplots}
\usepackage{pgfplotstable}
\usepackage{DejaVuSans}
\usepackage{amsthm}
\usepackage{hyperref} % optional
\usepackage[T1]{fontenc}
\pgfplotsset{compat=1.18}
\usepgfplotslibrary{groupplots, fillbetween, statistics}
\usepackage{titlesec}
\usepackage{tikz}
\usetikzlibrary{matrix, positioning, arrows, shapes.geometric, patterns, fit}
\usepackage{quantikz}
\def\BibTeX{{\rm B\kern-.05em{\sc i\kern-.025em b}\kern-.08em
    T\kern-.1667em\lower.7ex\hbox{E}\kern-.125emX}}

\begin{document}

\title{Rethinking How to Act: Action-Space Engineering for Reinforcement Learning-Based Circuit Routing in Distributed Quantum Systems
}

\author{\IEEEauthorblockN{Joost van Veen}
\IEEEauthorblockA{\textit{Quantum \& Computer Engineering} \\
\textit{Delft University of Technology}\\
Delft, The Netherlands \\
j.j.vanveen@student.tudelft.nl}
\and
\IEEEauthorblockN{Luise Prielinger}
\IEEEauthorblockA{\textit{Quantum \& Computer Engineering} \\
\textit{Delft University of Technology}\\
Delft, The Netherlands \\
l.p.prielinger@tudelft.nl}
\and
\IEEEauthorblockN{Sebastian Feld}
\IEEEauthorblockA{\textit{Quantum \& Computer Engineering} \\
\textit{Delft University of Technology}\\
Delft, The Netherlands \\
s.feld@tudelft.nl}
}

% \author{
% \IEEEauthorblockN{
% Joost van Veen\textsuperscript{1},
% Luise Prielinger\textsuperscript{*,1,2}, and
% Sebastian Feld\textsuperscript{1,2}
% }
% \IEEEauthorblockA{
% \textsuperscript{1}Quantum \& Computer Engineering, Delft University of Technology, The Netherlands\\
% \textsuperscript{2}QuTech, Delft University of Technology, The Netherlands\\
% \textsuperscript{*}Corresponding author, l.p.prielinger@tudelft.nl.
% }
% }

\maketitle
\thispagestyle{plain}
\pagestyle{plain}
\pagenumbering{arabic}
\begin{abstract}
As it becomes increasingly difficult to monolithically scale a quantum processor, distributed quantum computing (DQC) offers an alternative by distributing qubits across multiple smaller interconnected quantum processor modules. In such an architecture, the challenge of quantum circuit compilation shifts from placing and routing qubits within one module to placing, routing and using the qubits efficiently across modules. In order to optimize circuit execution time, the right state-dependent networking decisions must be found, such as when and where to generate shared remote quantum states to support remote operations. 

Reinforcement learning (RL) provides a natural framework for this problem, generating a compilation policy that can generalize across different circuits. Building on the framework of Promponas et al. (2024), we introduce an agent that combines a novel action-space formulation with effective action-masking strategies. A comprehensive numerical comparison of the two approaches under different coupling constraints shows that our agent achieves improved training and inference performance with a relative reduction in the modeled execution time of up to 35\%. 
%Finally, we carry out a critical assessment of both agents based on our results, lay out identified limitations and outline directions toward further improving scalability and applicability of RL-based compilation for DQC.
\end{abstract}


\section{Introduction}
\begin{figure}
    \centering
    \includegraphics[width=1.\linewidth]{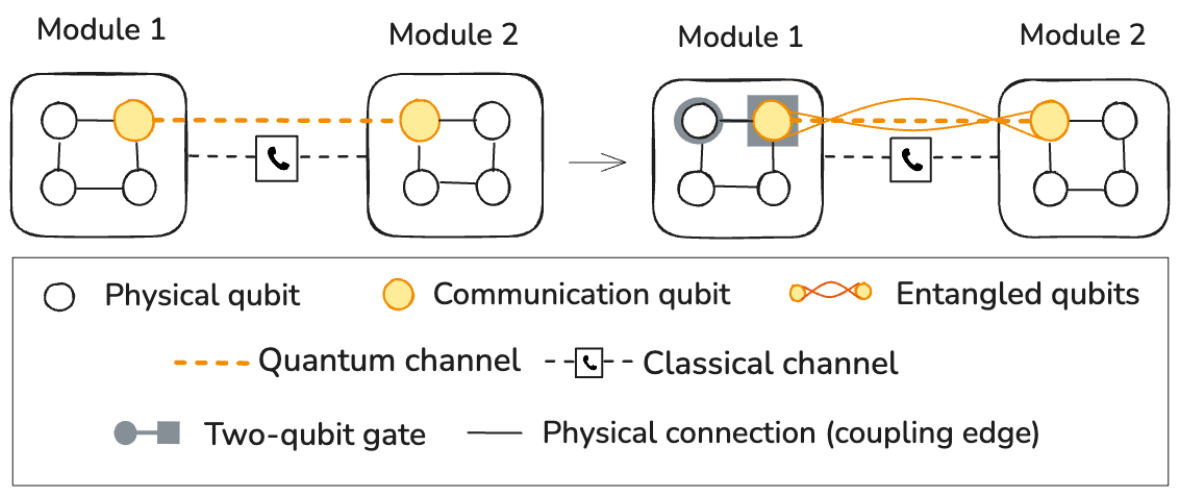}
    \caption{An entangled state is generated between two remote qubits used as communication qubits connected via a quantum channel and heralded by a classical message sent over a classical channel. Within each module, local two-qubit gates can be applied between all qubits according to the available connectivity (physical coupling edges).}
    \label{fig:overview}
\end{figure}

Sufficiently large, fault-tolerant quantum computers will have significant implications for various fields, ranging from fundamental physics~\cite{georgescu_quantum_2014}, over cryptography~\cite{shor_algorithms_1994} to chemistry and drug discovery~\cite{cao2018potential}. On today’s quantum computers, however, short-lived coherence times and imperfect control constrain computations to circuits that are still too shallow to be useful~\cite{preskill_quantum_2018,eisert2025mind}. We will need quantum error correction~\cite{terhal2015quantum}, encoding a logical qubit into \emph{many} physical qubits in order to make quantum computation fault-tolerant. However, monolithic scaling, i.e., adding ever more qubits to a single processor, is limited by the growing complexity of qubit control and calibration, as well as increased crosstalk and unwanted couplings that introduce correlated errors~\cite{de2021materials}. A promising alternative route to increasing the number of qubits in a single processor is interconnecting multiple smaller and therefore better controllable modules into a modular architecture~\cite{monroe_large-scale_2014,cirac1999distributed,main_distributed_2025, nickerson_topological_2013}, drawn in Fig.~\ref{fig:overview}. Here, quantum operations span networked quantum modules linked by quantum and classical communication channels. When shared inter-module quantum states are realized as maximally entangled qubit-pairs, i.e., Einstein-Podolsky-Rosen (EPR) pairs~\cite{einstein1935can}, then the rate at which they can be generated is typically much slower than the rate at which single- or two-qubit gates can be executed.
For example, in many solid-state platforms such as Nitrogen-vacancy centers~\cite{humphreys_deterministic_2018, kalb2017entanglement}, an heralded remote entangled state between two meter-spaced modules is established only at $10$Hz to $40$Hz, whereas a local single-qubit electron-spin rotation is set by microwave Rabi rates of order $10\,\mathrm{MHz}$, and electron-nuclear two-qubit gates are typically implemented with dynamical-decoupling sequences of a few kHz. 
% As a result, local gates might be several times faster than inter-QPU communication and the latter therefore takes up the larger share of end-to-end execution time in modular quantum systems. 
Thus, compared to monolithic compilation, distributed circuit compilation must not only satisfy local connectivity constraints within each module, but also decide how to realize non-local interactions through remote EPR pairs that are much more limited in their generation rate. 

Early work in distributed circuit compilation has largely focused on static qubit placement, i.e., the initial assignment of each virtual qubit in the circuit to a physical qubit in a module, and the minimization of inter-module interaction in order to reduce the number of entangled inter-module states. Approaches in this line typically split the quantum circuit into discrete time stages. At each stage, a so-called interaction graph is derived from concurrent multi-qubit gates, with an edges encoding pairwise interaction. The set of qubits is then partitioned into modules and assigned down to the physical qubits, which can be achieved by solving the graph-partitioning problem and (sub)graph isomorphism problem, respectively~\cite{andres2019automated, baker_time-sliced_2020, bandic2023mapping}. In these works, inter-module communication is typically collapsed into a single scalar cost of inter-module communication counts. Such an abstraction captures the amount of non-local operations, but no network dynamics. Advancing this line of research, several works develop network-aware methods that explicitly account for EPR-pair generation and routing of qubits within and between modules. For example, Cuomo et al.~\cite{cuomo_optimized_2023} formulate the routing task as a flow problem with capacity-limited entanglement links; a solution then selects, in each round, a maximal feasible set of tele-gates (those are gates that are executed between two modules; thereby consuming an EPR pair). Or, Ferrari et al.~\cite{ferrari_modular_2023} first partition the circuit, place each partition on a module, and then schedule remote operations, and finally compiling locally using conventional intra-module routing strategies, such as inserting SWAP gates along shortest paths~\cite{li2019tackling}. Most recent work further extends these approaches by integrating coherence times and probabilistic EPR-pair generation, thereby producing compilation decisions that pertain to the underlying network dynamics~\cite{sundaram_dqc-qr_2025}. However, they still rely on heuristics that may require long execution times to obtain good solutions, and this cost must be paid per circuit. 
%First, an initial placement maps each virtual qubit to a specific QPU and physical qubit, thereby determining which circuit gates must be executed remotely. The resulting set of required entangled pairs is then ordered by consumption and partitioned accordingly into batches for entanglement generation: In a batch of entangled pairs the compiler chooses, for each associated remote interaction, whether to use a tele-gate or a so-called cat-entanglement~\cite{nielsen_quantum_2011}. A cat-entanglement is a bipartite state that mirrors the state of a virtual qubit onto another QPU and can there be used as a control-type operand (``read-only'') for two-qubit gates until a single-qubit gate on the original qubit is executed; at that point the linked mirror-state is removed. Based on this decision, the algorithm determines the set of EPR pairs that must be generated. In the second stage, local QPU routing is performed while the EPR pairs for the next batch are being generated, so local routing and EPR-pair generation can proceed in parallel.

%These models incorporate scheduling of bell pairs into mapping and routing decisions produce more realistic and therefore more sensible solutions than earlier work focusing solely on the minimization of a highly abstracted inter-core communication.
As an effective alternative, reinforcement learning (RL) can be used to iteratively learn a strategy that naturally generalizes across multiple circuits seen during training. Depending on the model size and hyperparameters, training a neural network can be computationally expensive~\cite{goodfellow2016deep}, potentially even more so than executing the aforementioned heuristics. However, once trained, the cost of applying the learned policy to new circuits is typically much lower than using previously described approaches. When compiling many circuits, therefore, there will always be a regime in which the total time spent on training and compilation is lower than that of repeatedly invoking computationally intensive non-learning algorithms. 

Furthermore, reinforcement learning has already shown promising results in qubit routing in non-distributed architectures. For example, Tang et al.~\cite{tang_alpharouter_2024} introduce AlphaRouter, a hybrid approach that combines RL with tree search; and Pozzi et al. employ an agent without any auxiliary search~\cite{pozzi_using_2022}. In benchmark evaluations, both approaches outperform standard and state-of-the-art heuristic routers, including SABRE~\cite{li2019tackling}. In addition, RL has recently been studied for the first time for routing quantum circuits in a distributed IBM quantum architecture by Promponas et al.~\cite{promponas_compiler_2024}. In their work, the compilation task is formulated as a general Markov decision process and subsequently approximated in a first numerical implementation.

While this formulation provides a flexible framework for studying distributed quantum circuit routing with reinforcement learning, it raises the question of which design choices can be improved to increase performance. RL depends strongly on the structure of the action space, the efficiency of exploration, and the computational cost of training. These considerations motivate the search for formulations that preserve validity of the approach while improving its scalability. In this context, the present work introduces a novel action-space formulation, as well as action masking strategies and efficient Q-value approximation, with the goal of improving both training behavior and inference performance within the framework introduced by Promponas et al.~\cite{promponas_compiler_2024}. In our numerical assessment, we use the latter as baseline and show that our approach requires less wall-clock training time while achieving better inference performance in all considered settings. We base our analysis on a set of randomly generated quantum circuits varying gate count and executed on two hardware graphs that differ significantly in their connectivity. %We analyze both training and inference behavior, including the impact of different lookahead settings. 
 % Finally, we use these results to identify concrete current limitations and motivate sensible future research directions.

The remainder of this paper is organized as follows. Section~\ref{sec:model} describes the quantum circuit and system model. Section~\ref{sec:rlfordqc} introduces the reinforcement-learning framework. Section~\ref{sec:ourmodel} presents our approach in detail and Section~\ref{sec:eval} provides the numerical evaluation.

\section{Quantum circuit and system model}\label{sec:model}
A quantum computation system executes quantum logic as sequences of single- and multi-qubit operations, collectively referred to as quantum circuits. These circuits are often represented as directed acyclic graphs (DAGs). Fig.~\ref{fig:quant-dag} depicts an example of a quantum circuit represented as DAG. We adopt the representation of Promponas et al.~\cite{promponas_compiler_2024} and define a $\mathrm{DAG} := (V_g, E_g)$, where $V_g$ is the set of gates and $E_g$ encodes precedence constraints as directed edges between gates. More specifically, a directed edge $(u,v) \in E_g$ means that operation $v$ is constrained to occur after operation $u$ in every valid schedule. For example, in Fig.~\ref{fig:quant-dag}, the two operations (H on $q_0$) and (CNOT $q_0 \rightarrow q_1$) must be executed subsequently as they act on the same qubit $q_0$. On the other hand, the operations (H on $q_0$) and (X on $q_2$) form a so-called \emph{layer} in this example circuit, as they can be executed in parallel. More generally, a layer $\ell_d$ is a subset of gates $\ell_d \subseteq V_g$ such that for all distinct $u,v\in \ell_d$,
(i) the gates do not overlap, i.e., $\mathrm{supp}(u)\cap \mathrm{supp}(v)=\emptyset$, and
(ii) there is no dependency edge between them, i.e., $(u,v)\notin E_g$ and $(v,u)\notin E_g$. Equivalently, layers are non overlapping sets of gates that group gates at the same logical circuit depth $d$, so that all gates in a layer can be executed in parallel. We also refer to the first layer, where $d=1$, also as \emph{frontier}.

\input{Figures/dag}

To compile such a circuit description on a specific hardware platform, the circuit must satisfy the connectivity constraints of the target architecture. Conventionally, the hardware connectivity is represented by a coupling graph
\begin{smallblock} \begin{align}\label{eq:couplinggraph}
G = (V, E),\ \mathrm{with}\quad E = E_n \cup E_c,\ \mathrm{and\ } E_n \cap E_c = \emptyset,
\end{align}
\end{smallblock}
with $V$ denoting the set of physical qubits.  An edge in $E_c$ represents a channel between two remote qubits, see Fig.~\ref{fig:overview}. The edge set $E_n$ denotes the coupling connections in a module enabling local operations. We assume that all local operations take some constant duration to execute, which we denote $t_{\mathrm{local}}$.
We note that remote entanglement generation over a quantum channel is inherently stochastic: each attempt succeeds with probability $p_{\mathrm{gen}}<1$ and otherwise must be repeated. We adopt a standard abstraction~\cite{promponas_compiler_2024, talsma2024cd, humphreys_deterministic_2018} and treat entanglement generation as deterministic with a fixed latency $t_{\mathrm{gen}}$, approximating the mean waiting time of a repeat-until-success protocol, $t_{\mathrm{gen}} := t_0/p_{\mathrm{gen}}$, where $t_0$ is the duration of a single heralded entanglement generation attempt.

\subsection{Remote quantum logic}
Once an EPR pair has been established between two remote physical qubits, the hardware can realize non-local primitives. In particular, we assume the considered system supports (i) \emph{tele-qubits} (Definition~\ref{def:telequbit}), which teleport an arbitrary qubit state between modules, and (ii) \emph{tele-gates} (Definition~\ref{def:telegate}), which implement a non-local CNOT gate~\cite{nielsen_quantum_2010}. Figures~\ref{fig:dqc_epr_tele_grid}(c) and (d) visually explain the described primitives. Executing a tele-gate and tele-qubit are approximated to take the same constant time, denoted $t_{\mathrm{remote}}$.
\input{Figures/actions_new}

\newtheorem{definition}{Definition}[section]

\begin{definition}[Tele-qubit]\label{def:telequbit}
Let $q$ be a virtual qubit residing on a physical qubit in QPU~1. Let $(\mathrm{E1},\mathrm{E2})$ be an EPR pair shared between the communication qubits on QPU~1 and QPU~2. The following operation transfers the quantum state of $q$ to $\mathrm{E2}$ via:
1) Apply $\mathrm{CNOT}_{q\to \mathrm{E1}}$. 2) Apply a Hadamard gate $H$ to $q$. 3) Measure $q$ and $\mathrm{E1}$ in the computational ($Z$) basis, obtaining $m_1,m_2\in\{0,1\}$. 4) Send $(m_1,m_2)$ to QPU~2 over a classical channel. 5) On QPU~2, apply the Pauli correction $X^{m_2} Z^{m_1}$ to $\mathrm{E2}$.
\end{definition}

\begin{definition}[Tele-gate]\label{def:telegate}
Let $c$ (control) be a virtual  qubit residing on a physical qubit in QPU~1 and let $t$ (target) be a virtual  qubit residing on a physical qubit in QPU~2. Let $(\mathrm{E1},\mathrm{E2})$ be an EPR pair shared between the communication qubits on QPU~1 and QPU~2. The following operation implements a non-local CNOT from $c$ to $t$ via: 1) Apply $\mathrm{CNOT}_{c\to \mathrm{E1}}$. 2) Measure $\mathrm{E1}$ in the computational ($Z$) basis, obtaining $m\in\{0,1\}$. 3) Send $m$ to QPU~2 over a classical channel. 4) On QPU~2, apply the Pauli correction $X^{m}$ to $\mathrm{E2}$. 5) Apply $\mathrm{CNOT}_{\mathrm{E2}\to t}$. 6) Measure $\mathrm{E2}$ in the $X$ basis, obtaining $m'\in\{0,1\}$. 7) Send $m'$ to QPU~1 over a classical channel. 8) Apply the Pauli correction $Z^{n}$ to $c$.
\end{definition}

\section{Reinforcement Learning for DQC} \label{sec:rlfordqc}
In this section we (A) introduce important concepts of the used machine learning approach, according to standard notation~\cite{sutton_reinforcement_nodate} and (B) outline the baseline framework~\cite{promponas_compiler_2024}.
\subsection{Important deep learning concepts}
Reinforcement learning (RL) is a type of machine learning where an agent learns to achieve a goal by interacting with an environment through applying actions. Unlike supervised learning, where models learn from labeled datasets, RL learns from experience through trial and error. At each step, the agent observes the current state of the environment, selects an action and receives feedback in the form of a reward along with a new state. The agent's objective is to select actions which maximize the expected sum of future rewards over time. Formally, it keeps an estimate of the so-called Q-function $Q(s,a)$, which maps each state-action pair $(s,a)$ to the expected cumulative reward obtained when taking action $a$ in a state $s$ and following the current policy thereafter.
In environments with large state spaces, it becomes computationally infeasible to represent and update all Q-values. Instead, the Q-function is approximated $\tilde{Q}(s,a; \theta)$ using a multi-layered neural network (NN); the NN takes a state $s$ as input and returns an estimated Q-value for a given action $a$. Here, $\theta$ is the collection of all trainable parameters of the neural network, such as its weights and biases. In Deep Q-Networks (DQN) a separate target network is used as follows: for a transition $(s,a,r,s')$ from a current state $s$ to the next state $s'$ given an action $a$ with immediate scalar reward $r$, the target $y$ is
\begin{smallblock} 
\begin{align}
y &=
\begin{cases}
r, & \text{if } s' \text{ is terminal}, \\
r + \gamma \displaystyle\max_{a'} \tilde Q(s',a';\theta^-), & \text{otherwise},
\end{cases}
\end{align}
\end{smallblock}
where $\gamma \in [0,1)$ is the discount factor and $a'$ presents the next action.
This target network uses a delayed copy $\theta^-$ of the main Q-network's parameters and is updated at fixed intervals. Without this mechanism, the Q-values could become unstable because the network would be learning from targets that are themselves changing too quickly. To further stabilize training, DQN does not learn after each action, instead it stores its experience in a so-called replay buffer $\mathcal{D}$ and learns from a random set of samples of recent experiences. This reduces the correlation between consecutive experiences, which are typically more similar to one another than randomly sampled experiences from the replay buffer. Learning directly from temporally adjacent transitions, in fact, can cause updates to be inadequately influenced by short-term local patterns in the environment, which is undesirable. With these definitions, we can denote the loss function $L$ as the mean squared error between the current Q-value and the target:
\begin{smallblock} \begin{align} \label{eq:loss}
L(\theta)
  &= \mathbb{E}_{(s,a,r,s') \sim \mathcal{D}}
     \Big[ \big( y - \tilde Q(s,a;\theta) \big)^2 \Big],
\end{align}
\end{smallblock}
which the main Q-network aims to minimize by calculating gradients with respect to $\theta$. Double Deep Q-Networks (DDQN) is an extension of DQN, which uses the same two networks but the target value is computed differently. In DQN the target network chooses the next best action and calculates the associated reward:
\begin{smallblock} \begin{align}
y_{\text{DQN}} &= r + \gamma \max_{a'} Q(s', a'; \theta^-).
\end{align}
\end{smallblock}
Computing the maximum over noisy estimates tends to introduce an overestimation bias. In DDQN the next best action $\hat a$ is therefore chosen by the main network,
\begin{smallblock} \begin{align}
\hat a &= \arg\max_{a'} Q(s', a'; \theta),
\end{align}
\end{smallblock}
but its value is evaluated using the target network:
\begin{smallblock} \begin{align}
y_{\text{DDQN}} &= r + \gamma\, Q(s', \hat a; \theta^-).
\end{align}
\end{smallblock}

Simply put, the main network decides which action looks best, while the target network provides a more stable estimate of how good that action is.

As the final ingredient, RL uses epsilon-greedy exploration to balance exploration (trying new actions) and exploitation (choosing the best-known action). The parameter $\varepsilon\in\mathbb{R}$ with $\ 0 < \varepsilon \le1$ controls the probability of selecting a random action instead of the action with the highest predicted Q-value, i.e., 
\begin{smallblock} 
\begin{align}
\hat a &= \begin{cases}
    a' \sim \text{Unif}(\mathcal{A}_s),\ \text{with probability } \varepsilon\\
    \hat a^*,\ \text{with probability } 1-\varepsilon,
\end{cases}
\end{align}
\end{smallblock}
where $\mathcal{A}_s$ is the set of possible actions in state $s$, which is a subset of all actions $\mathcal{A}_s \subset \mathcal{A}$. At the start of training, where $\varepsilon = \varepsilon_0 = 1$, the agent choosing random actions. Over time, $\varepsilon$ decreases, and as such, the agent is relying more on its learned Q-values.

\subsection{DQC setting}
In this section we describe the reinforcement-learning cycle introduced by Promponas et al.~\cite{promponas_compiler_2024}, including its most important numerical building blocks, such as the action space $\mathcal{A}$, the state space $\mathcal{S}$, the Q-value approximation and the action masking, and lastly, how the reward is calculated. 

\paragraph{The action space $\mathcal{A}$}
\noindent We first distinguish between \emph{operations} supported by the compilation framework and \emph{actions} available to the agent. To execute a quantum circuit, the framework supports the following operations: the execution of a SWAP or CNOT gate on any local edge in $E_n$, and the generation of an EPR pair on any remote channel in $E_c$. If an EPR pair is available, the framework further supports the execution of a \emph{tele-gate} or a \emph{tele-qubit} operation on a remote connection in $E_c$; see Fig.~\ref{fig:dqc_epr_tele_grid} and definitions in Section~\ref{sec:model}. For simplicity, we neglect single-qubit gates, as these are purely local operations and therefore do not require routing nor remote communication. Among these operations, CNOT and tele-gate directly implement a circuit gate and therefore remove the corresponding vertex $v \in V_g$ from the DAG when executed. By contrast, SWAP, generate, and tele-qubit modify the placement of virtual qubits or prepare the resources required for remote execution. We therefore refer to CNOT and tele-gate as \emph{delete operations}, and to SWAP, generate, and tele-qubit as \emph{routing actions}. Importantly, as delete operations take priority whenever they are feasible, they are executed automatically by the environment and are therefore \emph{not} part of the agent's action space.

Each operation takes a set number of timesteps to be executed, summarized in Table \ref{tab:timeperoperation}. The listed values are in accordance with the assumptions of Promponas et al.~\cite{promponas_compiler_2024} and can be modified; however, they should reflect the condition that remote primitives (see Definitions \ref{def:telegate} and \ref{def:telequbit}) take several times longer than a local two-qubit operation. It can thus happen that an action cannot yet be executed due to required qubits being occupied. Furthermore, some actions might not be possible due to certain requirements not being met. For example, a tele-qubit can only be scheduled on a link in $E_c$ when an EPR pair already exists on this link. Invalid actions, i.e., actions not belonging to the admissible action set $\mathcal{A}_s$ in state $s$, need to be filtered out by a mask. 

\begin{table}[h]
    \centering
    \caption{Number of timesteps each action takes to complete~\cite{promponas_compiler_2024}.}
    \begin{tabular}{|c|c|}
        \hline
        Actions & Execution time \\ \hline
        Score ($t_{\mathrm{local}}$) & 1 \\
        Swap ($3 \times t_{\mathrm{local}}$) & 3 \\
        Generate ($t_{\mathrm{gen}}$)& 5 \\
        Tele-gate ($t_{\mathrm{remote}}$)& 5 \\
        Tele-qubit ($t_{\mathrm{remote}}$)& 5 \\ \hline
    \end{tabular}
    \label{tab:timeperoperation}
\end{table}

In addition to the actions listed in Table \ref{tab:timeperoperation}, the agent can wait for a timestep to go by when it is infeasible to perform any routing action. This action is termed STOP.
In a given state $s$ the agent introduced by Promponas et al. can therefore choose from the set of currently feasible actions $\mathcal{A}_s \subset \mathcal{A}$, where
\begin{smallblock}
\begin{align}
    \mathcal{A} = 
\left\{
\begin{aligned}
  &\mathrm{STOP},\, \mathrm{SWAP}_1,\ldots,\mathrm{SWAP}_{|E_n|},\\
  &\mathrm{telequbit}_1,\ldots,\mathrm{telequbit}_{|E_c|},\\
  &\mathrm{generate}_{1},\ldots,\mathrm{generate}_{|E_c|}
\end{aligned}
\right\}
\end{align},
\end{smallblock}
with size $|\mathcal{A}| = 1 + 2|E_c| + |E_n|$. 
\paragraph{The state space $\mathcal{S}$}
An environment's state $s\in\mathcal{S}$ is a vector that stores the qubit mapping and the $\mathrm{DAG}$ as they transform under the agent's actions. In particular, whenever a gate $v\in V_g$ of the $\mathrm{DAG}$ gets executed, the vertex $v$ is deleted and the $\mathrm{DAG}$ is updated, $V_g \leftarrow V_g\setminus{\{v\}}$ and $E_g \leftarrow E_g\setminus{E_v}$, where $E_v$ denotes all edges incident to $v$. Fig.~\ref{fig:state_vector}c) shows an example of the state vector, given a qubit mapping in Fig.~\ref{fig:state_vector}a) and a $\mathrm{DAG}$ in Fig.~\ref{fig:state_vector}b). The qubit-mapping is associated with the first $|V|$ indices of the state vector, where the $i$-th index indicates the physical qubit $i\in V$ and the value $s_i\in s$ refers either to the mapped virtual qubit or stores -1, if the qubit is not initialized (i.e., does not have a virtual qubit mapped to it). Entries at $(|V|+1)$ through $(|V|+3|V_g|)$ of the state vector store 3-tuples $v \mapsto (x,y,d)$, where $(x,y)$ denotes the ordered pair of virtual qubits involved in the gate, and $d$ denotes the layer $\ell_d$ of the $\mathrm{DAG}$ in which the gate appears.

\input{Figures/state_vector}

\paragraph{Reward structure}
An action's reward should reflect the agent's progress toward the overall goal: reducing execution time of the quantum circuit under the given hardware model. The agent in Promponas et al. receives a reward $R_{\text{score}}$ when an action leads to completing a gate, also referred to as ``scoring''. Successfully completing \emph{all} gates provides an additional reward, $R_{\text{success}}$. If the agent fails to complete the circuit, i.e., it cannot finish before a preset number of time steps $T$, it receives a large penalty, $R_{\text{fail}}$. Because the agent's goal is to route circuits as quickly as possible, also the STOP action is penalized. Each STOP action carries a small negative reward, $R_{\text{STOP}}$, such that minimizing unnecessary STOPs leads to a higher final reward. 

For qubit-movement actions, such as SWAP or tele-qubit operations the reward is based on the progress made toward routing the virtual qubits participating in frontier gates towards each other. In order to calculate the distance between locations of virtual qubits, the framework uses a distance metric~\cite{pozzi_using_2022}. Here, each state $s$ is associated with a weighted graph $G_s$, where physical couplings have weight $1$, quantum links have weight $w>1$, and the two halves of a generated entangled pair are connected by an edge of weight $1$. For every frontier gate $v\in\ell_1$, the shortest-path distance between the current locations of $x$ and $y$ is computed in $G_s$. Summing these distances over all frontier gates defines the state distance metric $d_{\mathrm{G}}(s)$. The reward of a movement action that transforms $s$ into $s'$ is then
\begin{smallblock} \begin{align} \label{eq:movereward}
R_{\mathrm{move}}(s,s')=\xi\bigl(d_{\mathrm{G}}(s)-d_{\mathrm{G}}(s')\bigr),
\end{align}
\end{smallblock}
where $\xi>0$ is a scaling constant. Thus, actions that reduce the total frontier distance receive positive reward.

\paragraph{The reinforcement-learning cycle}
The training of an RL agent is split into $N$ training cycles, so called \emph{episodes}. Each episode begins with a randomly generated initial mapping and quantum circuit, which together define the initial state $s^{\mathrm{init}}$. The state $s^{\mathrm{init}}$ is passed as input to the neural network. and the Q-values produced by the network are masked according to which actions are feasible in the current state $\mathcal{A}(s^{\mathrm{init}}) \subset \mathcal{A}$. At the start of training, the agent prioritizes exploration, as $\varepsilon_0 =  1$, selecting among the available actions randomly. The value of epsilon decays with each episode according to $\varepsilon(k) = \frac{\varepsilon_0}{1+\frac{k}{\varepsilon_d}}$, where $k \in \{0, 1,\dots,N\}$ is the episode number and $\varepsilon_d\in \mathbb{R^+}$ is a chosen epsilon decay denominator. After an action is chosen, any applicable delete gate $v \in \ell_{d=1}$ of the frontier layer of the DAG is executed and the latter is updated accordingly. When all gates of the frontier layer are executed, the agent proceeds to the next layer $\ell_{d=2}$. 

At last, the agent receives the reward and associated experience, which includes the state before $s$, the action chosen $a$ and the state after $s'$. This tuple $(s, a, s')$ is saved to the memory buffer $\mathcal{D}$. If the predefined number of steps between learning iterations has elapsed it performs the learning iterations, i.e., updates the agent's parameters via gradient descent using the loss function according to Equation (\ref{eq:loss}) and randomly sampling a batch of experiences of size $\beta$ from the replay buffer $\mathcal D$. Finally, the agent verifies whether the episode end criteria have been met, which is either the $\mathrm{DAG}$ being empty or the current time being passed a predefined number of maximum timesteps $T$. If the criteria are not yet met, the state and mask are updated, and the agent repeats the described steps until the end of the episode.

\section{Introduced Agent}\label{sec:ourmodel}
In this section, we describe the design choices for our agent, which is compatible with the framework introduced by Promponas at al., as they share the same state space $\mathcal{S}$. %We first introduce our action space $\tilde{\mathcal{A}}$ (\ref{sec:A}). Next, we introduce a restrictive masking strategy in \ref{sec:masking} that keeps only actions with high relevance. Finally, we detail our Q-value approximation in \ref{sec:qvals}, which reduces the number of trainable outputs, and we specify the resulting reward structure used during training in \ref{sec:reward}.
\subsection{Action space $\mathcal{\tilde A}$} \label{sec:A}
In our approach, instead of associating actions with edges in $E_n$, we allow an action to be composed of several SWAP and tele-qubit operations. More specifically, an action is associated with \emph{every} physical qubit pair $(i,j)$, where $\{i,j\} \subset V$ of the coupling graph (defined in Eq.~\ref{eq:couplinggraph}) and $i\neq j$. An action can rout the virtual qubit mapped to the first qubit $i$ in the pair to another physical qubit $j$ using a chain of SWAP and tele-qubit operations along a precomputed shortest path between $i$ and $j$ using Dijkstra's algorithm\cite{dljkstra1959note}. Note that in case of multiple shortest paths between two locations $(i,j)$, we select one at random. We refer to a routing operation as ROUT$_{(i,j)}$. 

Furthermore, we allow the STOP action to account for more than one time step. In particular, whenever a STOP action is implemented, time is advanced until at least one action becomes feasible again, we denote this elapsed time as $\Delta t_{\mathrm{skip}}$. The action space therefore becomes
\begin{smallblock}
\begin{align}
\tilde{\mathcal A}
=
\{\mathrm{STOP}\}
\cup
\left\{
\mathrm{ROUT}_{(u,v)}
\,\middle|\,
u,v \in V,\ u \neq v
\right\}
\cup
\left\{
\mathrm{gen}_{e}
\,\middle|\,
e \in E_c
\right\}.
\end{align}
\end{smallblock}
with size $|\mathcal{\tilde{A}}|=1 + |E_c| + |V|(|V|-1)$, where $|E_c|$ is the number of quantum links which connect different QPUs and $|V|$ is the number of physical qubits. Since $\mathcal{A} \subset \mathcal{\tilde{A}}$, our agent faces a larger degree of freedom with the potential of choosing more impactful actions at every step. However, this enlargement of the action space is naturally calling for a more restrictive action masking strategy in order to preserve the learning ability of the agent, which we will introduce in the following section.

\subsection{Action masking and Q-value approximation} \label{sec:masking}
In order to determine a restrictive action masking strategy, we define the admissible action set $\tilde A(s)\subset\tilde{A}$ in a state $s$ such that routing actions are retained only if they are directly related to progress in the current frontier or to the preparation of remote communication. Recall that the frontier is given by the first layer $\ell_1$ of the DAG. For a frontier gate $v\in \ell_1$, let $(x,y,1)$ denote the corresponding tuple in the state representation, where $x$ and $y$ are the two virtual qubits involved in that gate. We refer to a physical qubit as a \emph{frontier qubit} if it currently stores one of these virtual qubits. Furthermore, a value $-1$ means the qubit is non-initialized, and $E_1$ or $E_2$ refers to an \emph{EPR qubit}. With this terminology, the action mask excludes any routing action that neither decreases the distance in $G_s$ between the qubits of a frontier gate nor moves a non-initialized qubit towards a communication qubit associated with some link in $E_c$. Consequently, the admissible routing actions are restricted to the following three classes: i) actions $\mathrm{ROUT}(i,j)$ that move a frontier qubit from $i$ along a precomputed shortest path towards the physical qubit currently holding its frontier-gate partner; ii) actions $\mathrm{ROUT}(i,j)$ that move a non-initialized physical qubit from $i$ along a precomputed shortest path towards a physical qubit incident to a link in $E_c$, thereby preparing EPR-pair generation; iii) actions $\mathrm{ROUT}(i,j)$ that move an EPR qubit and a frontier qubit along precomputed shortest paths towards one another.
%In this way, the mask removes routing actions that are unlikely to contribute to near-term progress and restricts $\tilde A(s)$ to actions that are directly relevant to frontier-gate execution or EPR preparation.

In principle, the Q-network could output one value for every action in $\tilde{\mathcal A}$. However, this would require one Q-value for every directed routing action $(i,j)$, which scales quadratically in the number of physical qubits. To reduce the number of trainable outputs, we adopt a structured approximation in which the network predicts a set of induced Q-values, $\mathcal{Q}$, of size $|\mathcal{Q}| = 1 + |E_c| + |V|$: one for $\mathrm{STOP}$, one for each generate action, and one scalar $Q_i$ for each physical qubit $i \in V$. Under this construction, the number of trainable outputs is quadratically reduced in the number of physical qubits, as $|\tilde A| - |\mathcal{Q}| = |V|(|V|-2)$. We define an induced action value for each directed routing action from qubit $i$ to qubit $j$ 
\begin{smallblock} 
\begin{align}
    Q_{ij} := (1-\alpha)Q_i + \alpha Q_j,
    \qquad 0 < \alpha < 0.5.
\end{align}
\end{smallblock}
Since $1-\alpha > \alpha$, this construction preserves directionality, such that $Q_{ij} \neq Q_{ji}$. Intuitively, a large $Q_i$ increases the value of all routing actions originating from $i$, whereas the $\alpha Q_j$ term biases the choice among possible targets. This parameterization is less expressive than learning each $Q_{ij}$ independently, because all pairwise routing values must satisfy the additive structure above. In particular, it cannot represent arbitrary source-target interactions. Nevertheless, it provides an efficient low-dimensional approximation that reduces the number of routing-related outputs from $O(|V|^2)$ to $O(|V|)$, while still allowing the network to distinguish directed actions.
Moreover, for hardware graphs that are not simple paths, this construction yields a smaller parameterization than the one used in the baseline framework: the number of trainable Q-values scales as $1+|E_c|+|V|$ rather than $1+2|E_c|+|V|$ in the baseline approach.

\subsection{Reward structure} \label{sec:reward}
Our model alters the move reward $R_{\mathrm{move}}$ depending on the distance metric $d_{\mathrm{G}}(s)$ of a state $s$ given in Equation (\ref{eq:movereward}). As our restrictive action mask already filters out most actions which do not provide progress towards the goal, the agent does not receive a negative reward if the distance metric increases. Instead it simply receives zero reward. Thus the reward is defined as
$\tilde R_{\text{move}} := \max(\xi \,(d_{\text{G}}(s) - d_{\text{G}}(s'), 0)$
with $\xi\in \mathbb{R^+}$ denoting the reward's weighting factor. Further, while in Promponas et al. each STOP action carries a small negative reward, $R_{\text{STOP}}$, we reduce remaining occupation time by multiple time steps in one STOP action. The associated reward is therefore based on the number of time steps $\Delta t_{\mathrm{skip}}$ (Section \ref{sec:A}) the chain of operations take $\tilde R_{\text{STOP}} = R_{\text{STOP}}\cdot \Delta t_{\mathrm{skip}}$.

\section{Evaluation}\label{sec:eval}
In this section, we evaluate the performance of our agent and compare it with the DDQN-based agent of Promponas et al.~\cite{promponas_compiler_2024}. Our goal is to assess performance under different circuit lengths and hardware connectivity constraints. As our primary performance metric, we use the modeled circuit execution time $T$, i.e., the total number of time steps elapsed when an episode is successful. An episode is considered successful, when the agent achieves the full execution of the circuit before a set time limit $T\le T_\mathrm{max}$. If the time limit is reached, but not all gates of the circuit could be executed the training episode still terminates but is considered failed. We choose this metric because it provides a single and consistent measure across all actions. In contrast to surrogate metrics such as the number of remote gates or inter-module communications, the modeled execution time or elapsed time of execution also accounts for operations that do not directly implement circuit gates, including moving non-initialized qubits, generating EPR pairs, and routing qubits that store one half of an EPR pair. 
%This is particularly relevant in our setting, where compilation decisions couple local routing, remote communication, and waiting times, and where all of these contribute to the total modeled execution time of a compiled circuit. Moreover, the resulting execution time can be interpreted as a coarse proxy for the lifetime over which quantum states must remain coherent during execution, and therefore provides some intuition for the coherence requirements imposed by a compiled schedule.  
In addition to this primary metric, we report the wall-clock training time $T_{\mathrm{wall}}$ to capture the computational cost of learning over a set number of circuits. 

We consider two 32-qubit hardware topologies in order to study the effect of markedly different connectivity structures on performance: two connected QPUs of IBM Q Guadalupe and a $4\times4$ grid, shown in Figures~\ref{fig:qpu_mirrored} and~\ref{fig:two-grid-qpus}, respectively.

\input{Figures/2x16arch}
\input{Figures/2x16grid}

\subsection{Numerical setup}

To study how performance scales with problem size, we generate three training sets, each consisting of 250 circuits acting on 18 virtual qubits. The three sets differ in the number of CNOT gates per circuit, namely 30, 40, and 50; which we denote $\mathcal{C}_{30}, \mathcal{C}_{40}$ and $\mathcal{C}_{50}$, respectively.

The hyperparameters of the learning procedure are shared across agents and are listed in Table~\ref{tab:hyperparam} and other execution parameters used throughout the experiments are given in Table~\ref{tab:add_param}.

\begin{table}[h]
    \centering
    \caption{Hyperparameters used for training the DDQN agent.}
    \begin{tabular}{|l|r|}
    \hline
        Hyperparameter & Value \\ \hline
        Learning rate & 0.00001 \\
        Batch size ($\beta$) & 2560 \\
        Memory buffer size ($|\mathcal{D}|$) & 100000 \\
        Epsilon decay denominator ($\varepsilon_d$) & 80 \\
        Discount rate ($\gamma$) & 0.99 \\
        Target network update parameter ($\tau$) & 0.001 \\
        Learning frequency & Every 5 actions \\
        Number of learning iterations & 10 \\ 
        Optimizer & Adam~\cite{kingma2014adam} \\ \hline
    \end{tabular}
    \label{tab:hyperparam}
\end{table}

\begin{table}[h]
    \centering
    \caption{Execution parameter values.}
    \begin{tabular}{|l|r|}
    \hline
        Parameter & Value \\ \hline
        Time limit ($T_\mathrm{max}$) & 1500 \\
        Entanglement probability ($p_{gen}$) & 0.95 \\
        Distance weight ($w$) & 30 \\
        Distance multiplier ($\xi$) & 18 \\
        Reward stop ($R_{stop}$) & -20 \\
        Reward score ($R_{score}$) & 500 \\
        Reward empty dag ($R_{success}$) & 3000 \\
        Reward deadline ($R_{fail}$) & -3000 \\ \hline
    \end{tabular}
    \label{tab:add_param}
\end{table}

The two agents differ architecturally only in the treatment of qubit-pair actions, as described in Section~\ref{sec:ourmodel}. Both neural networks employ two hidden layers with ReLU activations. Promponas et al.~\cite{promponas_compiler_2024} use 150 and 140 neurons in the first and second hidden layers, respectively. In our experiments, we use the same neural network size for the agent trained on 50-gate circuits. For the 40-gate circuits a smaller second layer of 120 neurons, and for the 30-gate circuits 90 and 80 neurons for the first and second hidden layer have already proven sufficient.

All wall-clock measurements were obtained on a workstation equipped with an AMD Ryzen 7 5700G CPU and an NVIDIA RTX 3070 GPU.

\subsection{Training performance on $4\times4$ grid}
First, we assess training performance on a set of randomly generated circuits $\mathcal{C}_{30}$, where each $C\in \mathcal{C}_{30}$ holds 30 CNOT gates and $|\mathcal{C}_{30}|=250$. Both the baseline and our agent iterate through this set and try to compile one circuit instance per episode. The circuits are compiled under a high-connectivity topology, using two $4\times4$ grid QPUs, depicted in Fig.~\ref{fig:two-grid-qpus}. 
% Compared to a path graph, a grid offers multiple alternative paths between qubits, which generally makes routing more forgiving in the sense that suboptimal early routing decisions can be made up for more easily later in the process.

Fig.~\ref{fig:grid-comp} shows the modeled circuit execution time during one training run on the grid topology. Each data point represents the moving average over ten episodes. Overall, we observe that our model requires more episodes before achieving a meaningful reduction in execution time. Once trained, however, it attains slightly lower execution times than the baseline agent on this architecture.

The baseline agent exhibits a more gradual improvement throughout training, due to its routing actions being defined on individual edges, which makes it more likely to identify locally improving actions early on in a topology that offers many alternative paths. By contrast, the policy improvement of our model is much more abrupt, in particular between episodes 100 and 110. Our agent selects larger routing chains at once, but does so along a single predefined shortest path between two physical qubits. On a grid topology, this can be disadvantageous at the beginning of training, since multiple shortest paths typically exist between two locations. If routing chains are chosen along arbitrarily fixed shortest paths, different chains are more likely to intersect, which introduces waiting times. Once the agent learns to favor routing actions whose shortest paths interfere less with one another, the resulting policy achieves competitive, and ultimately even slightly better execution times.

\input{Figures/grid-comp}

\subsection{Training performance on Guadalupe}
\begin{figure}[h]
\centering
\includegraphics[width=\linewidth]{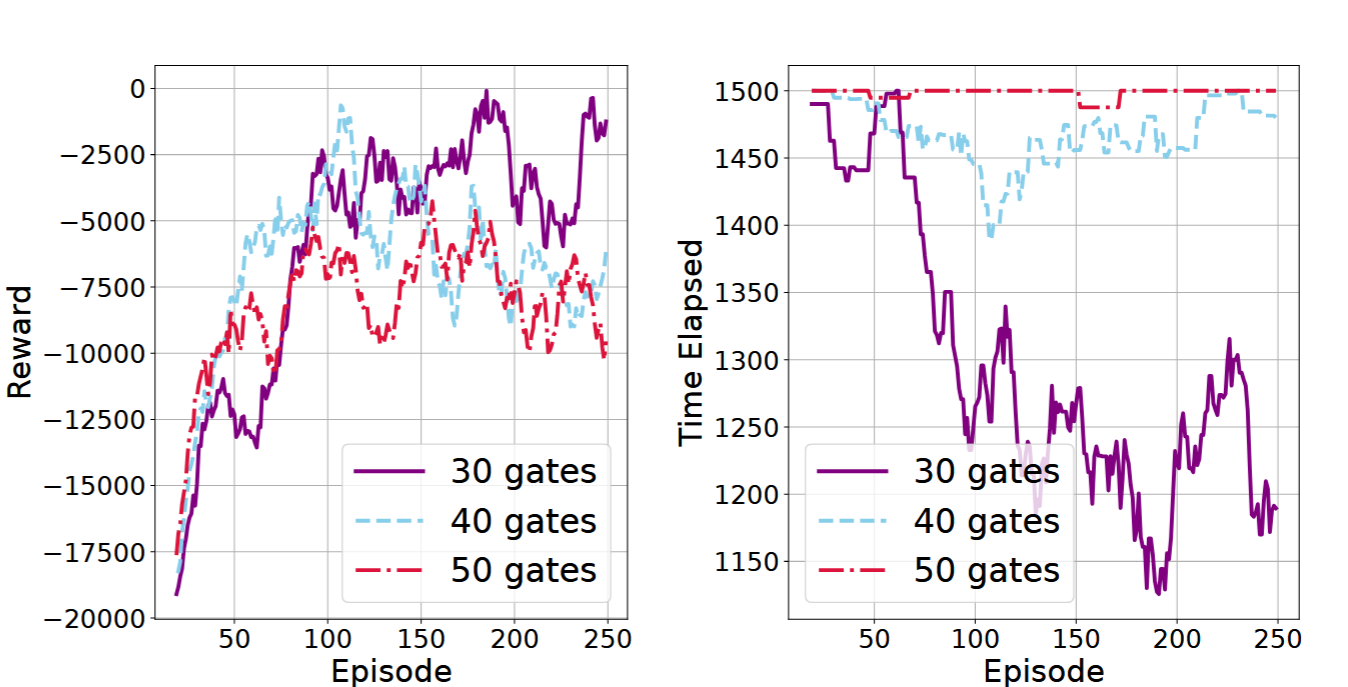}
\caption{Training results of baseline agent~\cite{promponas_compiler_2024} over 250 episodes of cumulative reward and execution time until successful quantum circuit compilation or deadline expiry for random circuits sets $\mathcal{C}_{30}, \mathcal{C}_{40}$ and $\mathcal{C}_{50}$. Each data point corresponds to the moving average of ten episodes.}
\label{fig:base-res}
\end{figure}

We first compare training performance on circuits with increasing gate count, $\mathcal{C}_{30}, \mathcal{C}_{40}$ and $\mathcal{C}_{50}$. Fig.~\ref{fig:base-res} and~\ref{fig:our_res} show the reward and execution time results reported in~\cite{promponas_compiler_2024} and our approach, respectively. Each datapoint corresponds to the moving average of ten episodes. As expected, the reward shows a large absolute improvement over training in the baseline approach; from $-$20 000 to $-$2 500 on average over the last 100 episodes. However, we observe that an increasing reward does not necessarily imply learning a policy that can do better than just randomly executing operations. We see in fact that for circuit sets $\mathcal{C}_{40}$ and $\mathcal{C}_{50}$ rewards increase but the execution times remain close to what can be achieved with randomly selecting and executing actions, i.e., no better-than-random policy is learned. Only for the set of circuits with smallest gate count $\mathcal{C}_{30}$, the baseline agent achieves a meaningful reduction in execution time to an average of $\sim$1210 over the last 100 episodes. In this setting, our model achieves lower execution time, depicted in Fig.~\ref{fig:our_res}. Over the last 100 episodes, the average solve time yields $\sim$746 time steps, corresponding to a relative improvement of $\sim$38\%. Furthermore and in contrast to the baseline agent, both agents trained on $\mathcal{C}_{40}$ and $\mathcal{C}_{50}$ exhibit a significant decrease in execution time toward the end of training, indicating improved scalability relative to the baseline. We see significantly less relative increase in our reward. Note that, however, we adapted our reward structure (see Section \ref{sec:reward}) and the reward results of baseline and our agent are therefore not directly comparable. 

We plot the standard deviation of the moving average in Fig.~\ref{fig:deviation_comp} to illustrate that both models exhibit variability throughout training. In general, large variation indicates that learning has not fully converged and remains sensitive to the specific circuit instances encountered. High reward variance further suggests that the learning signal is noisy or unstable, so that the agent does not achieve a consistent return across episodes. In this respect, our agent achieves a reduction in its standard deviation over the last 100 episodes, it attains an average standard deviation of approximately 2000, whereas the baseline agent remains more or less consistent above 4000 of its cumulative reward throughout training.

\input{Figures/circuit-size-comp}

\input{Figures/deviation}

The wall-clock runtime of training $T_{\mathrm{wall}}$ differs substantially. On 30-gate circuits, the agent of~\cite{promponas_compiler_2024} requires approximately 66 hours, whereas our model completes the set number of episodes in about 23.5 hours, a relative reduction of 64\%. We attribute this reduction primarily to the smaller neural network size as well as to the reduced number of Q-values.

\subsection{Inference performance and dependence on lookahead} \label{sec:eval_lookahead}
In this section, we study inference performance, i.e., how well an agent does on a separate test set of 250 random circuits after training is completed. More specifically, we evaluate and compare the agents trained on the $\mathcal{C}_{30}$ on a test set $\mathcal{C}^\mathrm{test}_{30}$ where each random circuit again contains 30 CNOT gates. We note that during training, after 250 episodes $\varepsilon(250) \approx 0.17$, so that the policy still includes exploration.

Fig.~\ref{fig:agents_performance_inference} shows the corresponding result distributions. As expected from the training results, the baseline agent remains at an overall larger execution time than our approach. Concretely, the average execution time of the baseline agent is approximately 1227, while our agent achieves an average execution time of approximately 799. These results are 2\% and 7\% higher relative to the training averages of the last 100 episodes, for the baseline and our agent, respectively. This correction is not surprising, as the observed shift remains small relative to the variability seen during training, and inference is after all evaluated on previously unseen test circuits. Still, the relative improvement in average execution time between introduced agent and baseline agent remains significant at $35$\%.

\input{Figures/inference_res}
Next, we assess whether it is necessary for an agent's inference performance to observe the \emph{full} circuit during training. A smaller lookahead would be desirable because it would reduce the size of the state representation and thus the computational burden of learning, potentially improving scalability to larger circuits. We therefore want to study to what extent restricting the visible part of the circuit affects inference performance. For this we evaluate our agent trained on $\mathcal{C}_{30}$, $\mathcal{C}_{40}$ and $\mathcal{C}_{50}$ on a test set $\mathcal{C}^\mathrm{test}_{50}$ with randomly generated circuits comprised of 50 CNOT gates, where $|\mathcal{C}^\mathrm{test}_{50}| = 250$. Fig.~\ref{fig:agent_performance_time} shows the associated distributions. We observe improvement from an agent trained on $\mathcal{C}_{30}$, where execution time average lies at approximately 1197 to an agent trained on $\mathcal{C}_{40}$, where the average execution time lies approximately at 1102, corresponding to about 7\% relative reduction; we do observe further improvement from the latter agent to the one trained on even larger circuits $\mathcal{C}_{50}$ where execution time average lies at approximately 995, another 9.7\% relative reduction. This suggests that in our setup, all involved gates impact routing decisions and training on larger circuits improves the agent's ability to make these decisions. We can therefore not assume that gates after a certain look-ahead window have less impact than earlier gates at the considered circuit size.
\input{Figures/partition_res}

\section{Conclusion}\label{sec:conclusion}
In this work, we investigated executing quantum circuits on distributed quantum computing (DQC) architectures using reinforcement learning. Aligned with the environment of an existing method by Promponas et al.~\cite{promponas_compiler_2024}, we introduced a novel action space, action masking strategies, as well as Q-value approximation and adapted the reward function. We compared our agent with the existing approach as baseline. The results show that both RL agents can learn nontrivial qubit-routing strategies for distributed quantum computing, and that the introduced model improves significantly over the baseline on the more constrained Guadalupe architecture while remaining competitive on a grid topology. Importantly, this improvement is achieved within the same state space. The main result is obtained for 30-gate circuits: during training, the average execution time decreases from an average over the last 100 episodes of approximately 1,210 timesteps for the baseline to approximately 746 timesteps for the proposed model (a reduction of about 38\%), see Figures \ref{fig:base-res} and \ref{fig:our_res}. This advantage also carries over to unseen test circuits, where inference execution time improves from about 1,227 to about 799 timesteps, corresponding to a relative improvement of approximately $35\%$ and depicted in Fig.~\ref{fig:agents_performance_inference}. On the more highly connected 4$\times$4 grid topology, the new model learns more slowly at the beginning of training, but it still reaches competitive final performance at the end of training. In addition, we tested the impact of look ahead windows, where we could see improvement in growing training size, with data presented in Fig.~\ref{fig:grid-comp}. Another important finding is that our proposed model is much less costly to train. In the same setting, when training on circuits of 30 gates relative wall-clock training time is reduced by 64\%, showing that the method is not only better performing but also more computationally efficient in the considered setting. However, scalability remains limited by the polynomial growth of the state space with the number of gates. Under the simplified modeling assumptions considered in this work, the agent assessments are still limited to 50 gates per circuit and 18 qubits under the current cost of training wall clock time. In addition, the masking strategy introduces a potential trade-off between faster learning and the ability to discover more globally optimized routing strategies. Future work will therefore focus on more compact and transferable state representations in order to support broader generalization and scalability. %On the modeling side, follow-up work should gradually relax the current abstractions by incorporating more realistic considerations, such as entanglement-generation scheduling, non-ideal quantum memory, and gate noise. Together, these directions will improve both the generalization ability and the practical applicability of the studied reinforcement learning for DQC approach.

\section{Data and code availability}
The complete dataset analyzed in this study is archived in
\cite{vanveen2026data}. The simulation code is part of
the same data archive, but is also available at \cite{vanveen_compilerdqc_2026}.
\section{Acknowledgments}
This work is supported by QuTech NWO funding 2020-2024 Part I “Fundamental Research”, Project Number 601.QT.001-1, financed by the Dutch Research Council (NWO). We further acknowledge support from NWO QSC grant BGR2 17.269. Generative AI disclosure: OpenAI’s ChatGPT Extended Thinking Model was used to generate plot skeletons into which the authors inserted numerically generated result data, and to generate TikZ code for selected Figures~\ref{fig:quant-dag}, \ref{fig:dqc_epr_tele_grid}, \ref{fig:state_vector}, \ref{fig:qpu_mirrored}, and \ref{fig:two-grid-qpus}.

{\footnotesize
\bibliography{references}}
\bibliographystyle{acm}

\end{document}

%% file: Figures/dag.tex
\begin{figure}[h]
    \centering
\resizebox{0.49\textwidth}{!}{%
    \begin{tikzpicture}[node distance=1cm,>=stealth,thick]
    \tikzstyle{gate} = [draw, rounded corners, minimum width=1.2cm, minimum height=0.8cm, align=center]
    \tikzstyle{layerlabel} = [font=\bfseries]
    
    % Nodes
    \node[gate] (H0) {H on $q_0$};
    \node[gate, below of=H0, yshift=-0.5cm] (X2) {X on $q_2$};
    
    \node[gate, right of=H0, yshift=-0.5cm, xshift=2cm] (CNOT01) {CNOT $q_0 \to q_1$};
    
    \node[gate, right of=CNOT01, yshift=-1.5cm, xshift=1cm] (X1) {X on $q_1$};
    
    \node[gate, above right=1cm and 1cm of X1] (Z0) {Z on $q_0$};
    % \node[gate, below of=Z0, yshift=-1.9cm, xshift=0.5cm] (CNOT12) {CNOT $q_1 \to q_2$};
    
    \node[gate, right of=Z0, yshift=-0.9cm, xshift=1cm] (CNOT01b) {CNOT $q_0 \to q_1$};
    
    % \node[gate, right of=CNOT01b, yshift=-1.5cm, xshift=1cm] (H1) {H on $q_1$};

    % Layer labels
    \node[layerlabel, above=0.01cm of H0]      {Layer 1};
    \node[layerlabel, above=0.01cm of X2]      {Layer 1};
    \node[layerlabel, above=0.01cm of CNOT01]  {Layer 2};
    \node[layerlabel, above=0.01cm of Z0]      {Layer 3};
    \node[layerlabel, above=0.01cm of X1]      {Layer 3};
    % \node[layerlabel, above=0.01cm of CNOT12]  {Layer 4};
    \node[layerlabel, above=0.01cm of CNOT01b] {Layer 4};
    % \node[layerlabel, above=0.01cm of H1]      {Layer 6};
    
    % Edges
    \draw[->] (H0) -- (CNOT01);
    \draw[->] (X2) -- (X1);
    
    \draw[->] (CNOT01) -- (X1);
    \draw[->] (X1) -- (CNOT01b);
    
    \draw[->] (CNOT01) -- (Z0);
    \draw[->] (Z0) -- (CNOT01b);
    
    %\draw[->] (CNOT12) -- (CNOT01b);
    
    % \draw[->] (CNOT01b) -- (H1);
    
    \end{tikzpicture}
    }
    \caption{DAG of a quantum circuit encoding precedence constraints of quantum operations. Operations in the same layer can be executed in parallel.}
    \label{fig:quant-dag}
\end{figure}  

%% file: Figures/actions_new.tex
\begin{figure}[t]
\centering
\footnotesize

\begin{subfigure}[t]{0.48\columnwidth}
\vspace{0pt}
\centering
\resizebox{\linewidth}{!}{%
\begin{tikzpicture}[
node/.style={circle, draw, thick, minimum size=6mm, inner sep=0.8pt, font=\footnotesize},
thickline/.style={line width=0.9pt},
epr/.style={red, thick, dashed},
telegate/.style={blue, ->, thick},
telequbit/.style={green, ->, thick},
highlight/.style={circle, draw, thick, minimum size=7mm, inner sep=0pt},
every node/.style={node}
]
\panelbboxtop

% Place physical-qubit annotations below for 2,3,6,7; above otherwise
\newcommand{\physpos}[1]{%
  \ifnum#1=2 below\else
  \ifnum#1=3 below\else
  \ifnum#1=6 below\else
  \ifnum#1=7 below\else
  above\fi\fi\fi\fi
}

% --- Left QPU ---
\foreach \n/\x/\y/\vlabel in {0/0/1/3, 1/1/1/-1, 2/0/0/0, 3/1/0/4} {
  \node[label=\physpos{\n}:{\scriptsize $\n$}] (\n) at (\x,\y) {\vlabel};
}
\foreach \a/\b in {0/1,1/3,3/2,2/0} {\draw[thickline] (\a) -- (\b);}

% --- Right QPU ---
\begin{scope}[xshift=3cm]
\foreach \n/\x/\y/\vlabel in {4/0/1/-1, 5/1/1/1, 6/0/0/2, 7/1/0/-1} {
  \node[label=\physpos{\n}:{\scriptsize $\n$}] (\n) at (\x,\y) {\vlabel};
}
\foreach \a/\b in {4/5,5/7,7/6,6/4} {\draw[thickline] (\a) -- (\b);}
\draw[thickline, dotted] ([xshift=-4cm]1) -- (4);
\end{scope}

\paneltagtop{a}
\end{tikzpicture}}
\end{subfigure}\hfill
\begin{subfigure}[t]{0.48\columnwidth}
\vspace{0pt}
\centering
\resizebox{\linewidth}{!}{%
\begin{tikzpicture}[
node/.style={circle, draw, thick, minimum size=6mm, inner sep=0.8pt, font=\footnotesize},
thickline/.style={line width=0.9pt},
epr/.style={red, thick, dashed},
telegate/.style={blue, ->, thick},
telequbit/.style={green, ->, thick},
highlight/.style={circle, draw, thick, minimum size=7mm, inner sep=0pt},
every node/.style={node}
]
\panelbboxtop

% --- Left QPU ---
\foreach \n/\x/\y/\label in {0/0/1/3, 1/1/1/-1, 2/0/0/0, 3/1/0/4} {
  \node (\n) at (\x,\y) {\label};
}
\foreach \a/\b in {0/1,1/3,3/2,2/0} {\draw[thickline] (\a) -- (\b);}

% --- Right QPU ---
\begin{scope}[xshift=3cm]
\foreach \n/\x/\y/\label in {4/0/1/-1, 5/1/1/1, 6/0/0/2, 7/1/0/-1} {
  \node (\n) at (\x,\y) {\label};
}
\foreach \a/\b in {4/5,5/7,7/6,6/4} {\draw[thickline] (\a) -- (\b);}
\draw[epr] ([xshift=-4cm]1) -- (4)
  node[midway, above, red, yshift=0.0cm, font=\footnotesize]{generate};
\node[highlight, red, fit=(1)] {};
\node[highlight, red, fit=(4)] {};
\end{scope}

\paneltagtop{b}
\end{tikzpicture}}
\end{subfigure}

\vspace{0.6em}

\begin{subfigure}[t]{0.48\columnwidth}
\vspace{0pt}
\centering
\resizebox{\linewidth}{!}{%
\begin{tikzpicture}[
node/.style={circle, draw, thick, minimum size=6mm, inner sep=0.8pt, font=\footnotesize},
thickline/.style={line width=0.9pt},
epr/.style={red, thick},
telegate/.style={blue, ->, thick},
telequbit/.style={green, ->, thick},
highlight/.style={circle, draw, thick, minimum size=7mm, inner sep=0pt},
every node/.style={node}
]
\panelbboxbottom

% Left QPU
\foreach \n/\x/\y/\label in {0/0/1/3, 1/1/1/E1, 2/0/0/0, 3/1/0/4} {
  \node (\n) at (\x,\y) {\label};
}
\foreach \a/\b in {0/1,1/3,3/2,2/0} {\draw[thickline] (\a) -- (\b);}

% Right QPU
\begin{scope}[xshift=3cm]
\foreach \n/\x/\y/\label in {4/0/1/E2, 5/1/1/1, 6/0/0/2, 7/1/0/-1} {
  \node (\n) at (\x,\y) {\label};
}
\foreach \a/\b in {4/5,5/7,7/6,6/4} {\draw[thickline] (\a) -- (\b);}
\draw[epr] ([xshift=-4cm]1) -- (4)
  node[midway, above, red, yshift=-1.2cm, font=\footnotesize]{EPR pair};
\node[highlight, red, fit=(1)] {};
\node[highlight, red, fit=(4)] {};
\end{scope}

\draw[telegate] (3) .. controls (1.5,2.5) .. ([xshift=3cm]5)
  node[midway, above, blue, font=\footnotesize]{tele-gate};
\node[highlight, blue, fit=(3)] {};
\node[highlight, blue, fit=(5)] {};

\paneltagbottom{c}
\end{tikzpicture}}
\end{subfigure}\hfill
\begin{subfigure}[t]{0.48\columnwidth}
\vspace{0pt}
\centering
\resizebox{\linewidth}{!}{%
\begin{tikzpicture}[
node/.style={circle, draw, thick, minimum size=6mm, inner sep=0.8pt, font=\footnotesize},
thickline/.style={line width=0.9pt},
epr/.style={red, thick},
telegate/.style={blue, ->, thick},
telequbit/.style={green, ->, thick},
highlight/.style={circle, draw, thick, minimum size=7mm, inner sep=0pt},
highlightgreen/.style={circle, draw, thick, minimum size=10mm, inner sep=1.5pt},
every node/.style={node}
]
\panelbboxbottom

% Left QPU
\foreach \n/\x/\y/\label in {0/0/1/3, 1/1/1/E1, 2/0/0/0, 3/1/0/4} {
  \node (\n) at (\x,\y) {\label};
}
\foreach \a/\b in {0/1,1/3,3/2,2/0} {\draw[thickline] (\a) -- (\b);}

% Right QPU
\begin{scope}[xshift=3cm]
\foreach \n/\x/\y/\label in {4/0/1/E2, 5/1/1/1, 6/0/0/2, 7/1/0/-1} {
  \node (\n) at (\x,\y) {\label};
}
\foreach \a/\b in {4/5,5/7,7/6,6/4} {\draw[thickline] (\a) -- (\b);}
\draw[epr] ([xshift=-4cm]1) -- (4)
  node[midway, above, red, yshift=-1.2cm, font=\footnotesize]{EPR pair};
\node[highlight, red, fit=(1)] {};
\node[highlight, red, fit=(4)] {};
\end{scope}

\draw[telequbit] (0) .. controls (1.5,3) .. ([xshift=3cm]4)
  node[midway, above, green, font=\footnotesize]{tele-qubit};
\node[highlight, green, fit=(0)] {};
\node[highlightgreen, green, fit=(4)] {};

\paneltagbottom{d}
\end{tikzpicture}}
\end{subfigure}

\caption{Panel (a) shows the qubit mapping: Annotated values $\{0,\dots,n-1\}$ denote physical qubit indices, while the value inside a node indicates the virtual qubit mapped onto the physical qubit; $-1$ implies a non-initialized physical qubit. In (b), the generate action creates an EPR pair on a pair of non-initialized remote communication qubits. Panel (c) visualizes a tele-gate execution and (d) visualizes a tele-qubit action.}
\label{fig:dqc_epr_tele_grid}
\end{figure}

%% file: Figures/state_vector.tex
\begin{figure}[h]
\centering

% --- (a) QPU mapping ---
\begin{subfigure}{0.95\linewidth}
\centering
\small
\begin{tikzpicture}[
    node/.style={circle, draw, thick, minimum size=6mm, inner sep=0.8pt, font=\footnotesize},
    thickline/.style={line width=0.9pt},
    every node/.style={node}
]

% --- Left QPU: 2x2 grid ---
\foreach \n/\x/\y/\label in {
    0/0/1/3,
    1/1/1/-1,
    2/0/0/0,
    3/1/0/4
} {
    \node (\n) at (\x,\y) {\label};
}
\foreach \a/\b in {0/1,1/3,3/2,2/0} {
    \draw[thickline] (\a) -- (\b);
}

% --- Right QPU: 2x2 grid ---
\begin{scope}[xshift=3cm]
    \foreach \n/\x/\y/\label in {
        4/0/1/-1,
        5/1/1/1,
        6/0/0/2,
        7/1/0/-1
    } {
        \node (\n) at (\x,\y) {\label};
    }
    \foreach \a/\b in {4/5,5/7,7/6,6/4} {
        \draw[thickline] (\a) -- (\b);
    }
    \draw[thickline, dotted] ([xshift=-4cm]1) -- (4);
\end{scope}
\end{tikzpicture}
\subcaption{Qubit mapping on a 2×2 QPU layout with one connecting quantum channel $|E_q| = 1$.}
\end{subfigure}

\par\medskip

% --- (b) DAG ---
\begin{subfigure}{0.95\linewidth}
\centering
\small
\begin{tikzpicture}[node distance=1.5cm,>=stealth,thick]
\tikzstyle{gate} = [draw, rounded corners, minimum width=2cm,
                    minimum height=0.9cm, align=left, font=\small]
\node[gate, label=above:{\small Gate 1}] (CNOT30) {CNOT $q_3 \to q_0$};

\node[gate, right of=CNOT30, xshift=1.5cm,
      label=above:{\small Gate 3}] (CNOT01) {CNOT $q_0 \to q_1$};

\node[gate, right of=CNOT01, xshift=1cm, yshift=0.0cm,
      label=above:{\small Gate 2}] (CNOT24) {CNOT $q_2 \to q_4$};

% Dependencies
\draw[->] (CNOT30) -- (CNOT01);

\end{tikzpicture}
\subcaption{DAG of a circuit, with frontier CNOT$(q_3, q_0)$, CNOT$(q_2, q_4)$.}
\end{subfigure}

\par\medskip

% --- (c) State vector ---
\begin{subfigure}{0.95\linewidth}
\small
\[
[\,3,\;-1,\;0,\;4,\;-1,\;1,\;2,\;-1,\;0,\;1,\;2,\;2,\;4,\;1,\;3,\;0,\;1\,]
\]
\begin{tabbing}
\hspace{0.5cm}\(\underbrace{\hspace{3.7cm}}_{\text{Mapping part}}\)
\(\underbrace{\hspace{3.6cm}}_{\text{DAG part}}\) \\
\hspace{4.2cm}\(\underbrace{\hspace{1.2cm}}_{\text{Gate 3}}\)
\(\underbrace{\hspace{1.2cm}}_{\text{Gate 2}}\)
\(\underbrace{\hspace{1.2cm}}_{\text{Gate 1}}\)
\end{tabbing}
\subcaption{Resulting state vector with the first $|V|$ entries describing the qubit mapping and the remaining entries encoding the DAG with Gates 1--3 using the 3-tuples $(0,1,2)$, $(2,4,1)$, $(3,0,1)$, respectively. Gate 1 and Gate 2 form the first layer $d=1$, i.e., the frontier, while Gate 3 is in the second layer $d=2$.}
\end{subfigure}

\caption{Example of a state vector.}
\label{fig:state_vector}
\end{figure}

%% file: Figures/2x16arch.tex
\begin{figure}[h!]
\centering
\footnotesize
\resizebox{\columnwidth}{!}{
\begin{tikzpicture}[
    node/.style={circle, draw, thick, minimum size=3.5mm, inner sep=0.8pt, font=\footnotesize},
    thickline/.style={line width=1.5pt},
    every node/.style={node}
]

% Left QPU node positions
\foreach \n/\x/\y in {
    0/6/3, 1/5/3, 2/5/2, 3/5/1, 4/4/3, 5/4/1,
    6/3/4, 7/3/3, 8/3/1, 9/3/0, 10/2/3, 11/2/1,
    12/1/3, 13/1/2, 14/1/1, 15/0/3
} {
    \node (\n) at (\x,\y) {\n};
}

% Left QPU edges
\foreach \a/\b in {1/2,2/3,3/5,5/8,8/11,11/14,14/13,13/12,12/10,10/7,7/4,4/1,6/7,0/1,8/9,12/15} {
    \draw[thickline] (\a) -- (\b);
}

% Right QPU (mirrored), shifted right
\begin{scope}[xshift=7.2cm]

    % Right QPU node positions (mirrored)
    \foreach \n/\x/\y in {
        16/6/3, 17/5/3, 18/5/2, 19/5/1, 20/4/3, 21/4/1,
        22/3/4, 23/3/3, 24/3/1, 25/3/0, 26/2/3, 27/2/1,
        28/1/3, 29/1/2, 30/1/1, 31/0/3
    } {
        \pgfmathtruncatemacro{\xmirrored}{6 - \x}
        \node (\n) at (\xmirrored,\y) {\n};
    }

    % Right QPU edges (same topology, offset node IDs by +16)
    \foreach \a/\b in {
        1/2, 2/3, 3/5, 5/8,
        8/11, 11/14, 14/13, 13/12,
        12/10, 10/7, 7/4, 4/1,
        6/7, 0/1, 8/9, 12/15
    } {
        \pgfmathtruncatemacro{\aR}{\a + 16}
        \pgfmathtruncatemacro{\bR}{\b + 16}
        \draw[thickline] (\aR) -- (\bR);
    }

    % Dotted line between left QPU node 0 and right QPU node 16
    \draw[thickline, dotted] ([xshift=-7.2cm]0) -- (16);

\end{scope}
\end{tikzpicture}
}
\caption{DQC system with two IBM Q Guadalupe QPUs. Node labels denote the physical qubit number; dotted line is the quantum channel connecting both modules.}
\label{fig:qpu_mirrored}
\end{figure}

%% file: Figures/2x16grid.tex
\begin{figure}[ht]
\centering
\resizebox{0.3\textwidth}{!}{%

    \begin{tikzpicture}[
      node/.style={circle, draw, thick, minimum size=5mm, inner sep=1pt, font=\scriptsize},
      thickline/.style={line width=2pt}
    ]

    % Scale to \columnwidth
    \begin{scope}[scale=0.85]

    % First 4x4 grid (left)
    \foreach \row in {0,...,3} {
      \foreach \col in {0,...,3} {
        \pgfmathtruncatemacro{\num}{\row*4 + (3 - \col)}
        \node[node] (L\num) at (\col,-\row) {\num};
      }
    }

    % Second 4x4 grid (right)
    \foreach \row in {0,...,3} {
      \foreach \col in {0,...,3} {
        \pgfmathtruncatemacro{\num}{16 + \row*4 + \col}
        \node[node] (R\num) at (6+\col,-\row) {\num};
      }
    }

    % Connections Left grid
    \foreach \row in {0,...,3} {
      \foreach \col in {0,...,2} {
        \pgfmathtruncatemacro{\numA}{\row*4 + (3 - \col)}
        \pgfmathtruncatemacro{\numB}{\row*4 + (3 - (\col + 1))}
        \draw[thickline] (L\numA) -- (L\numB);
      }
    }
    \foreach \col in {0,...,3} {
      \foreach \row in {0,...,2} {
        \pgfmathtruncatemacro{\numA}{\row*4 + (3 - \col)}
        \pgfmathtruncatemacro{\numB}{(\row+1)*4 + (3 - \col)}
        \draw[thickline] (L\numA) -- (L\numB);
      }
    }

    % Connections Right grid
    \foreach \row in {0,...,3} {
      \foreach \col in {0,...,2} {
        \pgfmathtruncatemacro{\numA}{16 + \row*4 + \col}
        \pgfmathtruncatemacro{\numB}{16 + \row*4 + \col + 1}
        \draw[thickline] (R\numA) -- (R\numB);
      }
    }
    \foreach \col in {0,...,3} {
      \foreach \row in {0,...,2} {
        \pgfmathtruncatemacro{\numA}{16 + \row*4 + \col}
        \pgfmathtruncatemacro{\numB}{16 + (\row+1)*4 + \col}
        \draw[thickline] (R\numA) -- (R\numB);
      }
    }

    % Connection between the two '0' nodes - dotted line
    \draw[thickline, dotted] (L0) -- (R16);

    \end{scope}

    \end{tikzpicture}
    }
    \caption{Qubit connectivity of DQC architecture consisting of two 4x4 grid modules, the dotted line indicates the quantum channel.}
    \label{fig:two-grid-qpus}
\end{figure}

%% file: Figures/grid-comp.tex
\begin{figure}[h]
    \centering
    \begin{tikzpicture}
    \begin{groupplot}[
        group style={
            group size=1 by 1,
            vertical sep=1.2cm,
        },
        width=0.95\columnwidth,
        height=5cm,
        tick label style={
            font=\sffamily\mdseries\tiny,
            /pgf/number format/assume math mode=true,
        },
        label style={font=\sffamily\mdseries\fontsize{6}{7}\selectfont},
        ymajorgrids=true,
        xmajorgrids=true,
        grid style={opacity=0.5},
        tick align=outside,
        xtick pos=bottom,
        ytick pos=left,
        xlabel shift={-6pt},
        ylabel shift={-6pt},
        xmin=0,
        xmax=160,
        trim axis left,
        tick style={major tick length=1pt},
        x tick label style={
            font=\sffamily\mdseries\tiny,
            yshift=2.5pt,
        },
        y tick label style={
            font=\sffamily\mdseries\tiny,
            xshift=2.5pt,
            /pgf/number format/.cd,
                use comma=false,
                1000 sep={},
        },
    ]

    % === Time - Mean Plot ===
    \nextgroupplot[
        xlabel={Episode},
        ylabel={Time Elapsed},
        xtick={50,100,...,250},
        ytick={300,400,...,1500},
        ymax=1600,
        ymin=200,
    ]
        \addplot[color=blue, thick] table [x=episode, y=mean, col sep=comma] {data/GridOld/doneTime_processed.csv};
        \addplot[color=red, thick] table [x=episode, y=mean, col sep=comma] {data/GridNew/doneTime_processed.csv};

        \addplot[name path=uppera, draw=none] table [
            x=episode, 
            y expr=\thisrow{mean}+\thisrow{std}, 
            col sep=comma
        ] {data/GridOld/doneTime_processed.csv};
        \addplot[name path=lowera, draw=none] table [
            x=episode, 
            y expr=\thisrow{mean}-\thisrow{std}, 
            col sep=comma
        ] {data/GridOld/doneTime_processed.csv};
        \addplot[blue!30, fill opacity=0.6] fill between[of=uppera and lowera];

        \addplot[name path=upperb, draw=none] table [
            x=episode, 
            y expr=\thisrow{mean}+\thisrow{std}, 
            col sep=comma
        ] {data/GridNew/doneTime_processed.csv};
        \addplot[name path=lowerb, draw=none] table [
            x=episode, 
            y expr=\thisrow{mean}-\thisrow{std}, 
            col sep=comma
        ] {data/GridNew/doneTime_processed.csv};
        \addplot[red!30, fill opacity=0.6] fill between[of=upperb and lowerb];

    % === Reward Plot ===
    % \nextgroupplot[
    %     xlabel={Episode},
    %     ylabel={Reward},
    %     scaled y ticks=false,
    %     yticklabel={\pgfmathprintnumber[int detect]{\tick}},
    %     xtick={50,100,...,250},
    %     ymax=25000,
    %     ymin=-25000,
    %     ytick={-20000,-15000,...,20000},
    % ]
    %     \addplot[color=blue, thick] table [x=episode, y=mean, col sep=comma] {data/GridOld/reward_processed.csv};
    %     \addplot[color=red, thick] table [x=episode, y=mean, col sep=comma] {data/GridNew/reward_processed.csv};

    %     \addplot[name path=uppera, draw=none] table [
    %         x=episode, 
    %         y expr=\thisrow{mean}+\thisrow{std}, 
    %         col sep=comma
    %     ] {data/GridOld/reward_processed.csv};
    %     \addplot[name path=lowera, draw=none] table [
    %         x=episode, 
    %         y expr=\thisrow{mean}-\thisrow{std}, 
    %         col sep=comma
    %     ] {data/GridOld/reward_processed.csv};
    %     \addplot[blue!30, fill opacity=0.6] fill between[of=uppera and lowera];

    %     \addplot[name path=upperb, draw=none] table [
    %         x=episode, 
    %         y expr=\thisrow{mean}+\thisrow{std}, 
    %         col sep=comma
    %     ] {data/GridNew/reward_processed.csv};
    %     \addplot[name path=lowerb, draw=none] table [
    %         x=episode, 
    %         y expr=\thisrow{mean}-\thisrow{std}, 
    %         col sep=comma
    %     ] {data/GridNew/reward_processed.csv};
    %     \addplot[red!30, fill opacity=0.6] fill between[of=upperb and lowerb];

    \end{groupplot}
    \end{tikzpicture}

    % Shared legend
    \vspace{0.5em}
    \begin{minipage}{\columnwidth}
    \centering
    \begin{tikzpicture}
    \begin{axis}[
        hide axis,
        xmin=0, xmax=1,
        ymin=0, ymax=1,
        legend columns=2,
        legend style={
            /tikz/every even column/.append style={column sep=0.5cm},
            draw=none,
            font=\sffamily\footnotesize,
            cells={anchor=west},
        }]
        \addlegendimage{no markers, blue}
        \addlegendentry{Promponas et al.~\cite{promponas_compiler_2024}}
        \addlegendimage{no markers, red}
        \addlegendentry{Our model}
    \end{axis}
    \end{tikzpicture}
    \end{minipage}

    \caption{Comparison of moving average of elapsed execution time $T$ durint training on 30-gate circuits in $\mathcal{C}_{30}$ compiled on two 4x4 grid QPUs. Error envelopes indicate the standard deviation of the moving average.}
    \label{fig:grid-comp}
\end{figure}

%% file: Figures/circuit-size-comp.tex
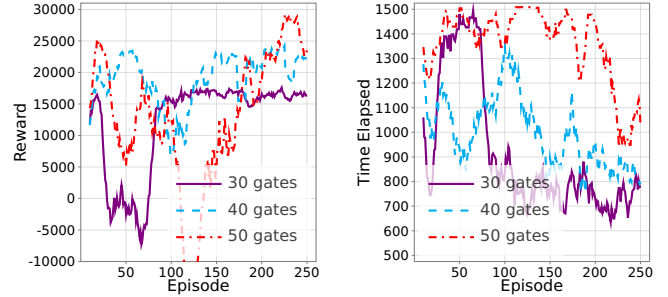
\begin{figure}[h]
    \begin{tikzpicture}
    \pgfplotsset{
        legend style={
            draw=none, 
            fill=white,             
            fill opacity=0.75,       
            font=\sffamily\mdseries\fontsize{6}{7}\selectfont,
        }
    }
    
    \begin{groupplot}[
        group style={
            group size=2 by 1,
            horizontal sep=1.3cm,
        },
        axis line style={draw=gray},
        width=0.53\linewidth,
        height=5cm,
        tick label style={
            font=\sffamily\mdseries\tiny,
            /pgf/number format/assume math mode=true,
        },
        label style={font=\sffamily\mdseries\fontsize{6}{7}\selectfont},
        ymajorgrids=true,
        xmajorgrids=true,
        grid style={opacity=0.5},
        tick align=outside,
        xtick pos=bottom,
        ytick pos=left,
        xlabel shift={-6pt},
        ylabel shift={-6pt},
        xmin=0,
        xmax=260,
        trim axis left,
        tick style={major tick length=1pt},
        x tick label style={
            font=\sffamily\mdseries\tiny,
            yshift=2.5pt,
        },
        y tick label style={
            font=\sffamily\mdseries\tiny,
            xshift=2.5pt,
            /pgf/number format/.cd,
                use comma=false,
                1000 sep={},
        },
    ]

    % --- Plot 1 (legend bottom left)
    \nextgroupplot[
        xlabel={Episode},
        ylabel={Reward},
        scaled y ticks=false,
        yticklabel={\pgfmathprintnumber[int detect]{\tick}},
        xtick={50,100,...,250},
        ymax=31000,
        ymin=-10000,
        ytick={-20000,-15000,...,30000},
        legend style={
            at={(0.98,0.02)},
            anchor=south east,
        },
    ]
        \addplot[color=violet, thick] table [x=episode, y=mean, col sep=comma] {data/30GatesNew/reward_processed.csv};
        \addplot[color=cyan, thick, dashed]  table [x=episode, y=mean, col sep=comma] {data/40GatesNew/reward_processed.csv};
        \addplot[color=red, thick, dashdotted] table [x=episode, y=mean, col sep=comma] {data/50GatesNew/reward_processed.csv};
        \legend{30 gates, 40 gates, 50 gates}
    
    % --- Plot 2 (legend bottom right)
    \nextgroupplot[
        xlabel={Episode},
        ylabel={Time Elapsed},
        xtick={50,100,...,250},
        ytick={500,600,...,1500},
        ymax=1525,
        ymin=475,
        legend style={
            at={(0.02,0.02)},
            anchor=south west,
        },
    ]
        \addplot[color=violet, thick] table [x=episode, y=mean, col sep=comma] {data/30GatesNew/doneTime_processed.csv};
        \addplot[color=cyan, thick, dashed]  table [x=episode, y=mean, col sep=comma] {data/40GatesNew/doneTime_processed.csv};
        \addplot[color=red, thick, dashdotted] table [x=episode, y=mean, col sep=comma] {data/50GatesNew/doneTime_processed.csv};
        \legend{30 gates, 40 gates, 50 gates}

    % % --- Plot 3: Time Std (bottom left)
    % \nextgroupplot[
    %     xlabel={Episode},
    %     ylabel={Time Elapsed},
    % ]
    %     \addplot[color=blue, thick] table [x=episode, y=std, col sep=comma] {data/30GatesNew/doneTime_processed.csv};
    %     \addplot[color=red, thick]  table [x=episode, y=std, col sep=comma] {data/40GatesNew/doneTime_processed.csv};
    %     \addplot[color=green, thick] table [x=episode, y=std, col sep=comma] {data/50GatesNew/doneTime_processed.csv};

    % % --- Plot 4: Reward Std (bottom right)
    % \nextgroupplot[
    %     xlabel={Episode},
    %     ylabel={Reward},
    %     yticklabel={\pgfmathprintnumber[int detect]{\tick}},
    %     scaled y ticks=false,
    % ]
    %     \addplot[color=blue, thick] table [x=episode, y=std, col sep=comma] {data/30GatesNew/reward_processed.csv};
    %     \addplot[color=red, thick]  table [x=episode, y=std, col sep=comma] {data/40GatesNew/reward_processed.csv};
    %     \addplot[color=green, thick] table [x=episode, y=std, col sep=comma] {data/50GatesNew/reward_processed.csv};

    \end{groupplot}
    \end{tikzpicture}

    % Shared legend
    % \vspace{0.5em}
    % \begin{center}
    % \begin{tikzpicture} 
    %     \begin{axis}[%
    %         hide axis,
    %         xmin=0, xmax=1,
    %         ymin=0, ymax=1,
    %         legend columns=3,
    %         legend style={
    %             /tikz/every even column/.append style={column sep=0.5cm},
    %             draw=none,
    %             font=\footnotesize,
    %             cells={anchor=west}
    %         }]
    %         \addlegendimage{no markers, purple}
    %         \addlegendentry{30 Gates}
    %         \addlegendimage{no markers, red}
    %         \addlegendentry{40 Gates}
    %         \addlegendimage{no markers, cyan}
    %         \addlegendentry{50 Gates};
    %     \end{axis}
    % \end{tikzpicture}
    % \end{center}

    \caption{Training results using our approach over 250 episodes of cumulative reward and execution time until successful quantum circuit compilation or deadline expiry for random circuits sets $\mathcal{C}_{30}, \mathcal{C}_{40}$ and $\mathcal{C}_{50}$. Each data point corresponds to the moving average of ten episodes.}
    \label{fig:our_res}
\end{figure}

%% file: Figures/deviation.tex
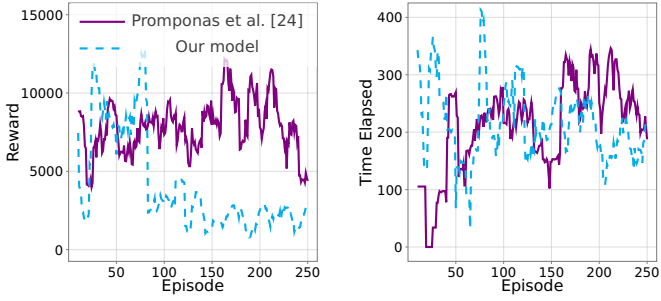
\begin{figure}[h]
    \begin{tikzpicture}
    \pgfplotsset{
        legend style={
            draw=none, 
            fill=white,             
            fill opacity=0.75,       
            font=\sffamily\mdseries\fontsize{6}{7}\selectfont,
        }
    }
    
    \begin{groupplot}[
        group style={
            group size=2 by 1,
            horizontal sep=1.2cm,
        },
        axis line style={draw=gray},
        width=0.55\linewidth,
        height=5cm,
        tick label style={
            font=\sffamily\mdseries\tiny,
            /pgf/number format/assume math mode=true,
        },
        label style={font=\sffamily\mdseries\fontsize{6}{7}\selectfont},
        ymajorgrids=true,
        xmajorgrids=true,
        grid style={opacity=0.5},
        tick align=outside,
        xtick pos=bottom,
        ytick pos=left,
        xlabel shift={-6pt},
        ylabel shift={-6pt},
        xmin=0,
        xmax=260,
        trim axis left,
        tick style={major tick length=1pt},
        x tick label style={
            font=\sffamily\mdseries\tiny,
            yshift=2.5pt,
        },
        y tick label style={
            font=\sffamily\mdseries\tiny,
            xshift=2.5pt,
            /pgf/number format/.cd,
                use comma=false,
                1000 sep={},
        },
    ]

    \nextgroupplot[
        xlabel={Episode},
        ylabel={Reward},
        scaled y ticks=false,
        yticklabel={\pgfmathprintnumber[int detect]{\tick}},
        xtick={50,100,...,250},
        ymax=15750,
        ymin=-750,
        ytick={0,5000,...,15000},
        legend style={
            at={(1.00,0.75)},
            anchor=south east,
        },
    ]
        \addplot[color=violet, thick] table [x=episode, y=std, col sep=comma] {data/30GatesOld/reward_processed.csv};
        \addplot[color=cyan, thick, dashed]  table [x=episode, y=std, col sep=comma] {data/30GatesNew/reward_processed.csv};
        \legend{Promponas et al.~\cite{promponas_compiler_2024}, Our model}
    
    % --- Plot 2 (legend bottom right)
    \nextgroupplot[
        xlabel={Episode},
        ylabel={Time Elapsed},
        xtick={50,100,...,250},
        ytick={0,100,...,400},
        ymax=425,
        ymin=-25,
        legend style={
            at={(0.3,0.02)},
            anchor=south west,
        },
    ]
        \addplot[color=violet, thick] table [x=episode, y=std, col sep=comma] {data/30GatesOld/doneTime_processed.csv};
        \addplot[color=cyan, thick, dashed]  table [x=episode, y=std, col sep=comma] {data/30GatesNew/doneTime_processed.csv};

    \end{groupplot}
    \end{tikzpicture}

    \caption{Standard deviation of training results of baseline~\cite{promponas_compiler_2024} and our model. Each datapoint reports standard deviation of the moving average over ten episodes for cumulative reward and execution time for circuit set $\mathcal{C}_{30}$.}
    \label{fig:deviation_comp}
\end{figure}

%% file: Figures/inference_res.tex
\begin{figure}[h]
\centering
\resizebox{0.7\columnwidth}{!}{%
\begin{tikzpicture}
    \begin{axis}[
        boxplot/draw direction=y,
        ylabel={Time Elapsed},
        xtick={1,2,3},
        xticklabels={
            Promponas et al.~\cite{promponas_compiler_2024},
            Our Model,
        },
        x tick label style={rotate=0, anchor=east},
        ymajorgrids,
        width=0.95\columnwidth,
        height=5cm,
        cycle list={},
        tick label style={
            font=\sffamily\mdseries\small,
            /pgf/number format/assume math mode=true,
        },
        label style={font=\sffamily\mdseries},
        ymajorgrids=true,
        xmajorgrids=true,
        grid style={opacity=0.5},
        tick align=outside,
        xtick pos=bottom,
        ytick pos=left,
        xlabel shift={-8pt},
        ylabel shift={-6pt},
        trim axis left,
        tick style={major tick length=1pt},
        x tick label style={
            font=\sffamily\mdseries,
            yshift=-10pt,
            rotate=0,
            xshift=0pt,
            anchor=center
        },
        y tick label style={
            font=\sffamily\mdseries,
            xshift=2.5pt,
            /pgf/number format/.cd,
                use comma=false,
                1000 sep={},
        },
        every boxplot/.style={
            fill=blue!20,
            draw=blue,
            every outlier/.style={draw=none,fill=none,mark=none}
        }
    ]
    \addplot[boxplot] table[y=time,col sep=comma] {data/30on30old/doneTime_processed.csv};
    \addplot[boxplot] table[y=time,col sep=comma] {data/30on30new/doneTime_processed.csv};
    \end{axis}
\end{tikzpicture}
}
\caption{Boxplots representing execution time comparison, where both baseline and our agent were trained on $\mathcal{C}_{30}$ executing the test circuits in $\mathcal{C}^{\mathrm{test}}_{30}$.}
\label{fig:agents_performance_inference}
\end{figure}
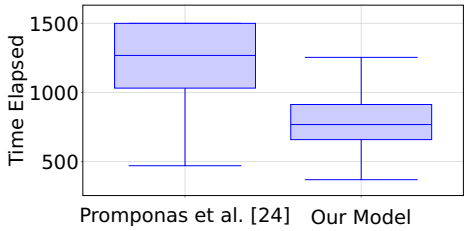

%% file: Figures/partition_res.tex
\begin{figure}[h]
\centering
\resizebox{0.7\columnwidth}{!}{%
\begin{tikzpicture}
    \begin{axis}[
        boxplot/draw direction=y,
        ylabel={Time Elapsed},
        xtick={1,2,3},
        xticklabels={
            $\mathcal{C}_{30}$,
            $\mathcal{C}_{40}$,
            $\mathcal{C}_{50}$
        },
        x tick label style={rotate=0, anchor=east},
        ymajorgrids,
        width=0.95\columnwidth,
        height=5cm,
        cycle list={},
        tick label style={
            font=\sffamily\mdseries,
            /pgf/number format/assume math mode=true,
        },
        label style={font=\sffamily\mdseries},
        ymajorgrids=true,
        xmajorgrids=true,
        grid style={opacity=0.5},
        tick align=outside,
        xtick pos=bottom,
        ytick pos=left,
        xlabel shift={-6pt},
        ylabel shift={-6pt},
        trim axis left,
        tick style={major tick length=1pt},
        x tick label style={
            font=\sffamily\mdseries,
            yshift=-10pt,
            rotate=0,
            anchor=center
        },
        y tick label style={
            font=\sffamily\mdseries,
            xshift=2.5pt,
            /pgf/number format/.cd,
                use comma=false,
                1000 sep={},
        },
        every boxplot/.style={
            fill=blue!20,
            draw=blue,
            every outlier/.style={draw=none,fill=none,mark=none}
        }
    ]
    \addplot[boxplot] table[y=time,col sep=comma] {data/30on50/doneTime_processed.csv};
    \addplot[boxplot] table[y=time,col sep=comma] {data/40on50/doneTime_processed.csv};
    % \addplot[boxplot] table[y=time,col sep=comma] {data/50on50/doneTime_processed.csv};
    \addplot[boxplot] table[y=time,col sep=comma] {data/50on50_2/doneTime_processed.csv};
    \end{axis}
\end{tikzpicture}}
\caption{Boxplots representing execution time results for our model trained on different circuits sizes ($\mathcal{C}_{30}$, $\mathcal{C}_{40}$ and $\mathcal{C}_{50}$) executing all circuits in $\mathcal{C}^{\mathrm{test}}_{50}$.}
\label{fig:agent_performance_time}
\end{figure}
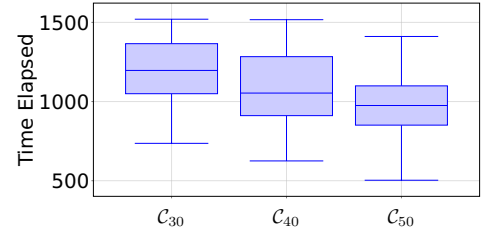